\PassOptionsToPackage{unicode}{hyperref}
\PassOptionsToPackage{hyphens}{url}

\PassOptionsToPackage{table}{xcolor}

\documentclass[
]{bio}

\usepackage{amsmath,amssymb}
\usepackage{lmodern}
\usepackage{iftex}
\ifPDFTeX
  \usepackage[T1]{fontenc}
  \usepackage[utf8]{inputenc}
  \usepackage{textcomp} 
\else 
  \usepackage{unicode-math}
  \defaultfontfeatures{Scale=MatchLowercase}
  \defaultfontfeatures[\rmfamily]{Ligatures=TeX,Scale=1}
\fi
\IfFileExists{upquote.sty}{\usepackage{upquote}}{}
\IfFileExists{microtype.sty}{
  \usepackage[]{microtype}
  \UseMicrotypeSet[protrusion]{basicmath} 
}{}
\makeatletter
\@ifundefined{KOMAClassName}{
  \IfFileExists{parskip.sty}{%
    \usepackage{parskip}
  }{
    \setlength{\parindent}{0pt}
    \setlength{\parskip}{6pt plus 2pt minus 1pt}}
}{
  \KOMAoptions{parskip=half}}
\makeatother
\usepackage{xcolor}
\providecommand{\tightlist}{%
  \setlength{\itemsep}{0pt}\setlength{\parskip}{0pt}}\usepackage{longtable,booktabs,array}
\usepackage{calc} 
\usepackage{etoolbox}
\makeatletter
\patchcmd\longtable{\par}{\if@noskipsec\mbox{}\fi\par}{}{}
\makeatother
\IfFileExists{footnotehyper.sty}{\usepackage{footnotehyper}}{\usepackage{footnote}}
\makesavenoteenv{longtable}
\usepackage{graphicx}
\makeatletter
\def\maxwidth{\ifdim\Gin@nat@width>\linewidth\linewidth\else\Gin@nat@width\fi}
\def\maxheight{\ifdim\Gin@nat@height>\textheight\textheight\else\Gin@nat@height\fi}
\makeatother
\setkeys{Gin}{width=\maxwidth,height=\maxheight,keepaspectratio}
\makeatletter
\def\fps@figure{htbp}
\makeatother
\newlength{\cslhangindent}
\newlength{\csllabelwidth}
\newlength{\cslentryspacingunit} 
\newenvironment{CSLReferences}[2] 
 {
  \setlength{\parindent}{0pt}
  \ifodd #1
  \let\oldpar\par
  \def\par{\hangindent=\cslhangindent\oldpar}
  \fi
  \setlength{\parskip}{#2\cslentryspacingunit}
 }%
 {}
\usepackage{calc}

\usepackage{setspace}
\makeatletter
\@ifpackageloaded{caption}{}{\usepackage{caption}}
\AtBeginDocument{%
\ifdefined\contentsname
  \renewcommand*\contentsname{Table of contents}
\else
  \newcommand\contentsname{Table of contents}
\fi
\ifdefined\listfigurename
  \renewcommand*\listfigurename{List of Figures}
\else
  \newcommand\listfigurename{List of Figures}
\fi
\ifdefined\listtablename
  \renewcommand*\listtablename{List of Tables}
\else
  \newcommand\listtablename{List of Tables}
\fi
\ifdefined\figurename
  \renewcommand*\figurename{Figure}
\else
  \newcommand\figurename{Figure}
\fi
\ifdefined\tablename
  \renewcommand*\tablename{Table}
\else
  \newcommand\tablename{Table}
\fi
}
\@ifpackageloaded{float}{}{\usepackage{float}}
\floatstyle{ruled}
\@ifundefined{c@chapter}{\newfloat{codelisting}{h}{lop}}{\newfloat{codelisting}{h}{lop}[chapter]}
\floatname{codelisting}{Listing}

\makeatother
\makeatletter
\@ifpackageloaded{caption}{}{\usepackage{caption}}
\@ifpackageloaded{subcaption}{}{\usepackage{subcaption}}
\makeatother
\ifLuaTeX
  \usepackage{selnolig}  
\fi
\IfFileExists{bookmark.sty}{\usepackage{bookmark}}{\usepackage{hyperref}}
\IfFileExists{xurl.sty}{\usepackage{xurl}}{} 
\hypersetup{
  pdftitle={Power and sample size calculations for testing the ratio of reproductive values in phylogenetic samples},
  pdfauthor={Lucy D'Agostino McGowan; Shirlee Wohl; Justin Lessler},
  hidelinks,
  pdfcreator={LaTeX via pandoc}}

\begin{document}
\begin{flushleft}
\centering
{\fontsize{17}{22}\bfseries{Power and sample size calculations for
testing the ratio of reproductive values in phylogenetic samples}\par}

\centering       
{\fontsize{10}{12}Lucy D'Agostino
McGowan\textsuperscript{1\dag*}, Shirlee
Wohl\textsuperscript{2,3}, Justin Lessler\textsuperscript{4,5,3\dag}}
\\
\bigskip
\textbf{1} Wake Forest University, Department of Statistical
Sciences, \\ \textbf{2} The Scripps Research Institute, Department of
Immunology and Microbiology, \\ \textbf{3} Johns Hopkins Bloomberg
School of Public Health, Department of Epidemiology, \\ \textbf{4} The
University of North Carolina at Chapel Hill, Department of
Epidemiology, \\ \textbf{5} The University of North Carolina at Chapel
Hill, Carolina Population Center, 
\bigskip

%
%
\dag These authors contributed equally to this work.



%

* mcgowald@wfu.edu

\end{flushleft}

\section*{Summary}
The quality of the inferences we make from pathogen sequence data is
determined by the number and composition of pathogen sequences that make
up the sample used to drive that inference. However, there remains
limited guidance on how to best structure and power studies when the end
goal is phylogenetic inference. One question that we can attempt to
answer with molecular data is whether some people are more likely to
transmit a pathogen than others. Here we present an estimator to
quantify differential transmission, as measured by the ratio of
reproductive numbers between people with different characteristics,
using transmission pairs linked by molecular data, along with a sample
size calculation for this estimator. We also provide extensions to our
method to correct for imperfect identification of transmission linked
pairs, overdispersion in the transmission process, and group imbalance.
We validate this method via simulation and provide tools to implement it
in an R package, phylosamp.
\doublespace

\hypertarget{introduction}{%
\section{Introduction}\label{introduction}}

Analysis of pathogen sequence data has revolutionized how we study
infectious disease, helping us address a myriad of questions ranging
from identifying the geographic origin and expansion of pathogens (Faria
\emph{and others} 2014), to reconstructing outbreaks (Gardy \emph{and
others} 2011; Wohl \emph{and others} 2020), to estimating the
transmissibility of emerging infectious diseases (Fraser \emph{and
others} 2009; Lai \emph{and others} 2020; Volz \emph{and others} 2020).
While many of the methods used to answer these questions are quite
sophisticated and leverage core principles of evolutionary and epidemic
theory (Stadler \emph{and others} 2012; MacPherson \emph{and others}
2022), much work is based on simply identifying transmission linked
pairs (Gardy \emph{and others} 2011; Ratmann \emph{and others} 2020).
Regardless of the sophistication of the methods used, the quality of the
inferences we make from pathogen sequence data is determined by the size
and structure of the sample of pathogen sequences used to drive that
inference. However, there remains limited guidance on how to best
structure and power studies when the end goal is phylogenetic inference.

In previous work we showed how sample size (i.e., the number of
sequences considered) and other study characteristics influenced the
identification of transmission linked pairs, including how many pairs a
study could expect to identify and the number of ``false positives''
(i.e., people deemed to be transmission linked that were not) that would
occur if the criteria used to define transmission linkage is imperfect.
However, while this work provided guidance on study design to identify
transmission linked pairs, it did not explore whether the resulting set
of pairs identified would be sufficient to answer an overarching
inferential question.

Here, we attempt to take that next step for one key question we might
ask from pathogen sequence data: whether some people are more likely to
transmit a pathogen than others. Specifically, we present an estimator
for the ratio of reproductive values between groups (i.e., individuals
with different characteristics) along with a sample size calculation for
this estimator. We explore the impact of imperfect observation of
transmission linked pairs (i.e., the sensitivity and specificity of the
linkage criteria) on required sample size, and address how other aspects
of the transmission process, such as super spreading and differential
susceptibility, impact this estimate.

\hypertarget{methods}{%
\section{Methods}\label{methods}}

\hypertarget{defining-transmission-pairs}{%
\subsection{Defining transmission
pairs}\label{defining-transmission-pairs}}

Often we are interested in asking whether there is differential
transmissibility of a pathogen between people with different levels of
some covariate, \(G\). For example, are older people more like to
transmit than younger people. A core measure of differences in
transmissibility is the ratio of the reproductive number of those with
covariate level `\(A\)' (\(G=A\)), denoted \(R_A\), versus those with
covariate level `\(B\)' (\(G=B\)), denoted \(R_B\). We assume the
characteristic of interest does not influence who you will be infected
by or who you will infect, however if may influence \emph{how many}
people you will infect (i.e.~transmission is not differential based upon
the relationship between the value of \(G\) of the infector and
infectee). For example, while we may be exploring whether men transmit
more than women, we are assuming that men are no more likely to infect
other men than women, and vice-versa.

\hypertarget{definitions-and-distributional-assumptions}{%
\subsection{Definitions and Distributional
Assumptions}\label{definitions-and-distributional-assumptions}}

We denote the size of the total infected population as \(N\) with
\(\pi_A\) representing the proportion for whom \(G=A\) and
\(\pi_B=1-\pi_A\). Hence, \(N_A = N\pi_A\) is the total infected
population for whom \(G=A\) and \(N_B = N\pi_B\) is the total infected
population in which \(G=B\). Each individual in the infected population
can be thought of as a \emph{node} connected to both their infector and
their infectees via \emph{edges}. Let us assume that our data consists
of pathogen sequences obtained from a simple random sample of infected
individuals such that the sampling proportion is \(\rho\) and the total
number of nodes sampled with each covariate level is as follows:
\(M_A = N_A\rho\), \(M_B = N_B\rho\).

A sampled transmission pair is defined as two individuals \(i\) and
\(j\) who are both included in the sample and are connected by an edge
(i.e., \(i\) infected \(j\)). If \(e_{i\cdot}\) is the total number of
infections in the sample caused by infector \(i\), then the total number
of directed edges in a sample of size \(M\) is \(E\):

\[E = \sum_{i=1}^M e_{i\cdot}\]

The probability that a transmission partner for infector \(i\) is
included in the sample is \(\frac{M-1}{N-1}\) (we subtract 1 from the
numerator and denominator to account for the infector) where \(M\) is
the total number of nodes in the sample. We assume the total number
infected by \(i\) (\(e_{i\cdot}\)) is Poisson distributed with a rate
\(\lambda = \frac{M-1}{N-1}R_{g_i}\)(Wohl \emph{and others} 2021) where
\(R_{g_i}\) is the total number of infectees for a given sampled
infection \(i\) belonging to group \(G=g\). Assuming \(e_{i\cdot}\) is
Poisson distributed, the sum of all \(e_{i\cdot}\) (\(E\)) is also
Poisson distributed with rate \(M\times \frac{M-1}{N-1}R_{pop}\). Where
\(R_{pop}\) is the average reproductive number for the infected
population (i.e.~\(R_{pop}\) is the total number of transmissions
divided by the total number of infected individuals, See Appendix A for
further intuition).

If we could perfectly observe transmission events, we would be able to
observe the frequency of four types of directed edges in our sample:

\begin{enumerate}
\def\labelenumi{(\arabic{enumi})}
\tightlist
\item
  \(E_{AA}\): The total number of directed edges in a sample where an
  individual in group \(A\) infects another individual in group \(A\)
\item
  \(E_{AB}\): The total number of directed edges in a sample where an
  individual in group \(A\) infects an individual in group \(B\)
\item
  \(E_{BA}\): The total number of directed edges in a sample where an
  individual in group \(B\) infects an individual in group \(A\)
\item
  \(E_{BB}\): The total number of directed edges in a sample where an
  individual in group \(B\) infects another individual in group \(B\)
\end{enumerate}

If these transmission pairs are identified using phylogenetic linkage
criteria (Wohl \emph{and others} 2021), such as the genetic distance
between infecting viruses, we often will not know who is the infector
and who is the infectee absent additional epidemiological information
(e.g.~the timing of symptom onset) or deep sequencing data.(Rose
\emph{and others} 2019) Hence we do not know the directionality of edges
and cannot differentiate \(E_{AB}\) from \(E_{BA}\) without further
assumptions, therefore we do not know the group of the infector for such
edges. However, for concordant edges in a sample, \(E_{gg}\) for
\(g\in\{A,B\}\), we do know the group of the infector, hence such edges
contain information about the group's relative likelihood of
transmitting the pathogen.

\hypertarget{defining-an-estimator-for-the-ratio-of-reproductive-numbers}{%
\subsection{Defining an Estimator for the Ratio of Reproductive
Numbers}\label{defining-an-estimator-for-the-ratio-of-reproductive-numbers}}

Using the definitions and distributional assumptions above, we define an
estimator for differential transmission as characterized by the ratio of
reproductive numbers between groups and define our hypothesis test in
relation to this estimator.

We assume the total number of observable group \(g\) concordant edges in
a sample of size \(M\) (edges where both the infector and infectee are
included in the sample) is Poisson distributed with a mean of
\(\lambda_g\), \(E_{gg}\sim Pois(\lambda_g)\) where

\begin{equation}\protect\hypertarget{eq-lambda}{}{
\lambda_g = \frac{M(M-1)R_g\pi_g^2}{(N-1)}
}\label{eq-lambda}\end{equation}

for \(g\in\{A,B\}\). See Appendix A for the derivation.

We are interested in testing whether the ratio of \(R\) values in each
group is equal to 1 versus greater than 1, in other words we conducting
the following one-sided hypothesis test:

\(H_0: R_A=R_B\) against \(H_1: R_B/R_A > 1\).

Equivalently, we are testing the null hypothesis that
\(\log(R_B/R_A)=0\), therefore we can develop a test statistic similar
to that specified in (Ng and Tang 2005),

Note that the sample estimate for \(R_g\) is

\begin{equation}\protect\hypertarget{eq-rhat}{}{
\hat{R}_g = \frac{E_{gg}(N-1)}{M(M-1)\pi_g^2}, g\in\{A,B\}.
}\label{eq-rhat}\end{equation}

Hence \(\frac{\hat{R}_B}{\hat{R}_A}\) is

\begin{equation}\protect\hypertarget{eq-rhatrat}{}{
\frac{\hat{R}_B}{\hat{R}_A} = \frac{E_{BB}\pi_A^2}{E_{AA}\pi_B^2}.
}\label{eq-rhatrat}\end{equation}

Giving us the following test statistic

\begin{equation}\protect\hypertarget{eq-u}{}{
U = \log(\hat{R}_B/\hat{R}_A)=\log(E_{BB}/E_{AA})-\log(d^2),
}\label{eq-u}\end{equation}

where \(d = \pi_B/\pi_A\). If unknown, since this is a simple random
sample, we can estimate this by \(d = M_B / M_A\) since
\(\hat\pi_g = M_g/M\) for \(g\in\{A,B\}\). Applying the delta method,

\begin{equation}\protect\hypertarget{eq-varu}{}{
\textrm{var}(U)=\sigma^2_U=\frac{(N-1)}{R_A\pi_A^2M(M-1)}+\frac{(N-1)}{R_B\pi_B^2M(M-1)}.
}\label{eq-varu}\end{equation}

This can be estimated by

\begin{equation}\protect\hypertarget{eq-su}{}{
s^2_U=\frac{(N-1)}{\tilde{R}_A\pi_A^2M(M-1)}+\frac{(N-1)}{\tilde{R}_B\pi_B^2M(M-1)},
}\label{eq-su}\end{equation}

where \(\tilde{R}_g\) is any reasonable estimate for \(R_g\) for
\(g\in\{A,B\}\). By letting \(\tilde{R}_g=\hat{R}_g\) for
\(g\in\{A,B\}\), we can use \(U/s_U\) to test the null hypothesis,
resulting in the following statistic,

\begin{equation}\protect\hypertarget{eq-w}{}{
W = \frac{\log(E_{BB}/E_{AA})-\log(d^2)}{\sqrt{1/E_{AA} + 1/E_{BB}}}.
}\label{eq-w}\end{equation}

Under the null hypothesis, \(W\) is asymptotically normally distributed,
with a mean of 0 and a standard deviation of 1, thus we reject the null
hypothesis when \(W>z_{1-\alpha}\) where \(z_{1-\alpha}\) is the
\(100\times(1-\alpha)\)th percentile of the standard normal
distribution. Note that this statistic does not exist when
\(E_{gg} = 0\).

\hypertarget{power-and-sample-size}{%
\subsection{Power and Sample Size}\label{power-and-sample-size}}

Now that we have an estimator, we can calculate the sample size needed
to detect pre-specified effects with a given power.

We can represent the variance
\((N-1)/R_A\pi_A^2M(M-1)+(N-1)/R_B\pi_B^2M(M-1)\) as
\(\left[1 + \frac{R_A}{d^2R_B}\right]\frac{(N-1)}{R_A\pi_A^2M(M-1)}\)
where \(d = \pi_B/\pi_A\) Under \(H_1: R_{B}/R_{A}>1\), we know
asymptotically \(\log(E_{BB}/E_{AA})-\log(d^2)\) follows a normal
distribution, as expressed below.

\[
\log(E_{BB}/E_{AA})-\log(d^2)\sim N\left(\log(R_{B}/R_{A}), \left[1 + \frac{R_A}{d^2R_B}\right]\frac{(N-1)}{R_A\pi_A^2M(M-1)}\right)
\]

Therefore the power can be expressed as

\begin{equation}\protect\hypertarget{eq-power}{}{
P=1-\Phi\left[z_{1-\alpha}-\frac{\log\left(\frac{R_B}{R_A}\right)}{\sqrt{\left[1 + \frac{R_A}{d^2R_B}\right]\frac{(N-1)}{R_A\pi_A^2M(M-1)}}}\right].
}\label{eq-power}\end{equation}

Hence, the total sample size needed to achieve \(1-\beta\) power under
\(H_1:R_B/R_A>1\) is

\begin{equation}\protect\hypertarget{eq-samp}{}{
M=\frac{1}{2}\left[1 + \sqrt{4\left[1+\frac{R_A}{R_Bd^2}\right]\frac{(z_{1-\alpha}+z_{1-\beta})^2}{\log(R_B/R_A)^2}\frac{(N-1)}{R_A\pi_A^2}+1}\right],
}\label{eq-samp}\end{equation}

where \(z_{1-\beta}\) is the \(100(1-\beta)\)th percentile of the
standard normal distribution.

\hypertarget{adjusting-for-sensitivity-and-specificity-of-linkage-criteria}{%
\subsection{Adjusting for sensitivity and specificity of linkage
criteria}\label{adjusting-for-sensitivity-and-specificity-of-linkage-criteria}}

When identifying transmission pairs using some phylogentic criteria, or
a combination of epidemiologic and phylogenetic criteria, the
transmission pairs actually identified in our sample will be dependent
on the sensitivity and specificity of that criteria (Wohl \emph{and
others} 2021). That is, since any criteria for identifying transmission
pairs will be imperfect, we will sometimes misclassify pairs of
individuals as being part of a transmission pair when they are not (and
vice versa). This matters because the expected number of observed
transmission pairs (edges) under misclassification, \(E^{*}_{gg}\),
differs from the expected number of sampled edges in the absence of
misclassification, \(E_{gg}\). The expected number of observed edges
under misclassification is

\[E^{*}_{gg}=E_{gg}(\eta + \gamma+1)+\frac{M_g(M_g-1)}{2} (1-\gamma), g\in\{A,B\}\]
where \(\eta\) is the assumed sensitivity level and \(\gamma\) is the
assumed specificity level.

Therefore, accounting for imperfect sensitivity and specificity, the
sample estimate for \(R_g\) is as follows,

\[\hat{R}_g = \frac{{E}^{*}_{gg} - \frac{M_{g}(M_g - 1)}{2}(1 - \gamma)}{\eta + \gamma - 1}\times\frac{(N-1)}{M(M-1)\pi_g^2}, g\in\{A,B\}.\]
First, let's consider the case where specificity is 100\%, and we have
imperfect sensitivity. Here, the number of edges observed will be
reduced from it true value, but out estimator will still be unbiased,
hence we can adjust our sample size estimate as follows

\begin{equation}\protect\hypertarget{eq-samp-sens}{}{
M=\frac{1}{2}\left[1 + \sqrt{4\left[1+\frac{R_A}{R_Bd^2}\right]\frac{(z_{1-\alpha}+z_{1-\beta})^2}{\log(R_B/R_A)^2}\frac{N-1}{R_A\eta\pi_A^2}+1}\right]
}\label{eq-samp-sens}\end{equation}

Where \(\eta\) is the estimated sensitivity. However, if specificity is
less than 100\% then our observed estimate of \(R_B/R_A\) will be biased
towards the null (assuming errors are non-differential). This bias will
further decrease our power. If \(\pi_A < \pi_B\) then the following will
be a conservative estimate for the sample size needed to achieve a given
power

\begin{equation}\protect\hypertarget{eq-samp-sens-spec}{}{
M=\frac{1}{2}\left[1 + \sqrt{4\left[1+\frac{R^*_A}{R^*_Bd^2}\right]\frac{(z_{1-\alpha}+z_{1-\beta})^2}{\log(R^*_B/R^*_A)^2}\frac{N-1}{R^*_A\pi_A^2}+1}\right]
}\label{eq-samp-sens-spec}\end{equation}

Where \(R^*_A = R_A(\eta + \gamma - 1) + (1 - \gamma) (N - 1)/2\),
\(R^*_B = R_B(\eta + \gamma - 1) + (1 - \gamma) (N - 1)/2\), and
\(\eta\) is the assumed sensitivity level and \(\gamma\) is the assumed
specificity level. Alternatively, if a more accurate estimate is
required, or \(\pi_B < \pi_A\), then the appropriate sample size can be
calculated by solving the transcendental equation that occurs when you
replace \(R^*_g\) with
\(R^*_g = R_g \times(\eta+\gamma-1)+ \frac{((M\pi_g - 1)/2)\times(1 - \gamma))(N-1)}{(M-1)\pi_g}\)
in Eq~\ref{eq-samp-sens-spec} (see the Appendix B for sample code and
implementation in the \texttt{phylosamp} package).

\hypertarget{correcting-for-overdispersion}{%
\subsection{Correcting for
overdispersion}\label{correcting-for-overdispersion}}

In real disease systems, the number of people each infected individual
infects is known not to be perfectly Poisson distributed. That is, the
presence of ``super-spreading events'' leads to overdispersion in the
transmission process.

To account for this, instead of assuming \(E_{gg}\sim Pois(\lambda_g)\)
as specified in Eq~\ref{eq-lambda}, which assumes the mean and variance
of the distribution are equivalent, we can instead allow \(E_{gg}\) to
follow a negative binomial distribution with an overdispersion parameter
\(k\). \(E_{gg}\) following a negative binomial distribution will change
our variance from Eq~\ref{eq-varu} to the following by the delta method

\begin{equation}\protect\hypertarget{eq-varu-negbin}{}{
\textrm{var}(U)=\sigma^2_U=\frac{(N-1)}{R_A\pi_A^2M(M-1)}+\frac{(N-1)}{R_B\pi_B^2M(M-1)}+\frac{2}{k}.
}\label{eq-varu-negbin}\end{equation}

This updates the power calculation as follows

\begin{equation}\protect\hypertarget{eq-power-k}{}{
P_{\textrm{overdisp}}=1-\Phi\left[z_{1-\alpha}-\frac{\log\left(\frac{R_B}{R_A}\right)}{\sqrt{\left[1 + \frac{R_A}{d^2R_B}\right]\frac{(N-1)}{R_A\pi_A^2M(M-1)}+\frac{2}{k}}}\right].
}\label{eq-power-k}\end{equation}

Therefore, the total sample size needed to achieve \(1-\beta\) power
under \(H_1:R_B/R_A> 1\) given the overdispersion parameter, \(k\) is

\begin{equation}\protect\hypertarget{eq-samp-k}{}{
M=\frac{1}{2}\left[1 + \sqrt{4\left[1+\frac{R_A}{R_Bd^2}\right]\frac{1}{\log(R_B/R_A)^2 / (z_{1-\alpha}+z_{1-\beta})^2 - 2/k}\frac{(N-1)}{R_A\pi_A^2}+1}\right].
}\label{eq-samp-k}\end{equation}

\hypertarget{two-sided-hypothesis-tests}{%
\subsection{Two-sided hypothesis
tests}\label{two-sided-hypothesis-tests}}

We can extend the above to the two-sided scenario where
\(H_0: R_A = R_B\) and \(H_1: R_B/R_A \neq 1\) as follows, where the
power for the two-sided test is

\begin{equation}\protect\hypertarget{eq-power-2}{}{
P_{\textrm{two sided}}=1-\Phi\left[z_{1-\alpha/2}-\frac{\left|\log\left(\frac{R_B}{R_A}\right)\right|}{\sqrt{\left[1 + \frac{R_A}{d^2R_B}\right]\frac{(N-1)}{R_A\pi_A^2M(M-1)}}}\right].
}\label{eq-power-2}\end{equation}

Therefore, the total sample size needed to achieve \(1-\beta\) power
under \(H_1:R_B/R_A\neq 1\) is

\begin{equation}\protect\hypertarget{eq-samp-2}{}{
M=\frac{1}{2}\left[1 + \sqrt{4\left[1+\frac{R_A}{R_Bd^2}\right]\frac{(z_{1-\alpha/2}+z_{1-\beta})^2}{\log(R_B/R_A)^2}\frac{(N-1)}{R_A\pi_A^2}+1}\right].
}\label{eq-samp-2}\end{equation}

\hypertarget{sample-size-correction}{%
\subsection{Sample size correction}\label{sample-size-correction}}

The asymptotic properties of these methods have been shown to
overestimate the needed sample size when there is imbalance between
groups such that \(\pi_B>\pi_A\) and underestimate when
\(\pi_B < \pi_A\) (Ng and Tang 2005; Gu \emph{and others} 2008). To this
end, we have implemented a simulation based approach that provides a
corrected sample size for a given desired power under specified
parameters. Functions to implement these simulations are available in
the \texttt{phylosamp} R package and sample code is provided in Appendix
B (Wohl \emph{and others} 2023).

\hypertarget{validation-methods}{%
\section{Validation Methods}\label{validation-methods}}

We validate our approach using two simulation frameworks. In the first,
``theoretical'' approach, we assume our distributional assumptions about
observed edges are correct and explore a wide range of the parameters
space. In the second, ``applied'' approach, we directly simulate
outbreaks and then sample from them. In this latter approach we explore
a smaller range of the parameter space due to computational constraints.

\hypertarget{theoretical-simulations}{%
\subsection{Theoretical Simulations}\label{theoretical-simulations}}

With the goal of covering a reasonable range of transmission difference
and group imbalance, we examine 31 values for the ratio \(R_B/R_A\) as
follows from 1 to 4 by 0.1 and 5 values for the proportion of the
population in the numerator, \(\pi_B = 0.05, 0.2, 0.5, 0.95\), resulting
in 155 scenarios. We simulate 100,000 ``epidemics'' under each scenario.
All epidemics have \(N = 100,000\) infected individuals (a large number
was selected to ensure achievable sample sizes). In each generated
epidemic, we simulated the total number of like edges in each group,
\(E_{AA}\) and \(E_BB\), by sampling from a Poisson distribution with
\(\lambda_g\) as specified in Eq~\ref{eq-lambda} based on the
specifications of each scenario. We perform a one-sided hypothesis test
testing the alternative hypothesis \(H_1: R_B/R_A > 1\). For scenarios
where \(R_B/R_A>1\) \(M\) is specified using the Eq~\ref{eq-samp},
setting the nominal power to 80\% (\(\beta=0.2\)) and \(\alpha = 0.05\).
Additionally, we calculate a ``corrected'' sample size,
\(M_{corrected}\) using a simulation based approach. For scenarios where
\(R_B/R_A = 1\) we sample \(\rho = 0.1\) to \(0.9\) proportion of the
epidemic. We then use the generated \(E_{AA}\) and \(E_BB\) values to
calculate the test statistic specified in Eq~\ref{eq-w} to perform the
one-sided hypothesis test in order to estimate the Type 1 (when
\(R_B/R_A = 1\)) and Type 2 (otherwise) errors.

Additionally, we consider scenarios where there is imperfect sensitivity
and specificity in our identification of transmission linked case pairs.
We examine sensitivity levels of 0.5, 0.75, 0.9 and 0.99 and specificity
levels of 0.99, 0.999, and 0.9999.

\hypertarget{applied-simulations}{%
\subsection{Applied Simulations}\label{applied-simulations}}

\hypertarget{generating-the-outbreaks}{%
\subsubsection{Generating the
outbreaks}\label{generating-the-outbreaks}}

We consider a wide range of potential ratios of \(R_B/R_A\) (1, 1.25,
1.5, 2, 2.5, 3, 3.5, 4) and 3 values for the proportion of the
population in the numerator, \(\pi_B\) = 0.2, 0.5, 0.8, resulting in 24
scenarios. We simulate 200 outbreaks under each scenario. In each
generated outbreak, we simulate the total number of susceptible
individuals, \(S_0\) from a Poisson distribution with the mean,
\(\lambda_s\), selected such that each scenario would generate outbreaks
of size \(N = 5,000\), on average. Each individual was assigned a group,
\(A\) or \(B\), with probability \(\pi_A\) and \(\pi_B\) as specified.
Simulated individuals moved through three stages: S (susceptible), I
(infectious), R (recovered). We randomly selected one individual to seed
the epidemic. The infected individual infected remaining susceptible
individuals with probability \(p_g = R_{0g}/ S_0\), where \(R_{0g}\) is
the basic reproductive number for group \(G=g\). Infected individuals
were infectious for one time step (i.e.~the epidemic progresses in
discrete generations). The epidemic continued until there were no
susceptible individuals remaining.

\hypertarget{sampling-from-the-outbreaks}{%
\subsubsection{Sampling from the
outbreaks}\label{sampling-from-the-outbreaks}}

We perform a one-sided hypothesis test testing the alternative
hypothesis \(H_1: R_B/R_A > 1\). For scenarios where \(R_B/R_A >1\), we
use Eq~\ref{eq-samp} to calculate the sample size, setting the nominal
power to 80\% (\(\beta=0.2\)) and \(\alpha = 0.05\). Additionally, we
calculate a ``corrected'' sample size using a simulation based approach.
For scenarios where \(R_B/R_A = 1\) sample \(\rho = 0.1\) to \(0.9\)
proportion of the outbreak. Using the specified sample size, \(M\) (and
additionally \(M_{corrected}\) for the scenarios where \(R_B/R_A>1\))
for each scenario, we randomly sample 10,000 times for each outbreak and
calculate the test statistic specified in Eq~\ref{eq-w} in order to
estimate the Type 1 and Type 2 errors.

\hypertarget{results}{%
\section{Results}\label{results}}

Over multiple simulation studies we find that the derived test
statistics and sample size calculations have the expected performance
properties.

\begin{figure}

{\centering \includegraphics{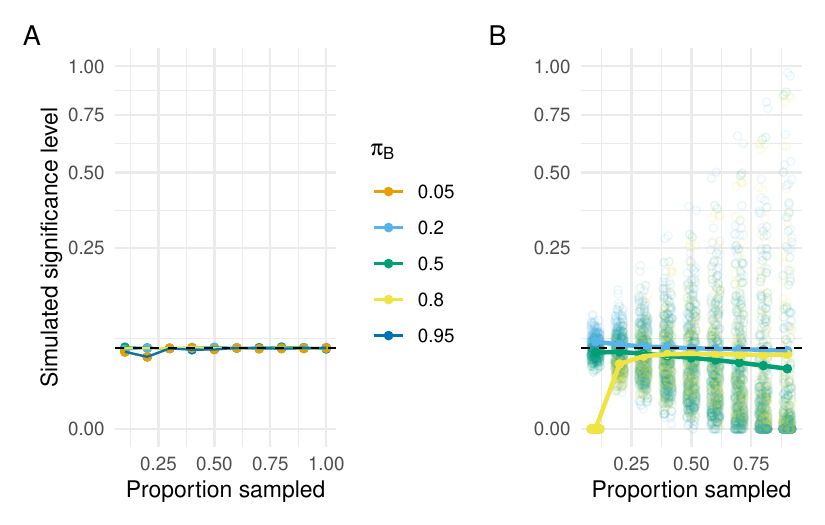}

}

\caption{\label{fig-null}Simulated type 1 error (\(\alpha = 0.05\),
shown in the dashed black line) varying the sampling fraction, \(\rho\),
from 0 to 0.9 (x-axis) \(\pi_B=0.05\) (orange, only in the theoretical
simulation), 0.2 (light blue), 0.5 (green), 0.8 (yellow), 0.95 (dark
blue, only in the theoretical simulation). A. Observed type 1 error in
the theoretical simulation B. Observed type 1 error in outbreak
simulation.}

\end{figure}

When there is no difference in transmissibility between groups, our
approach successfully bounds the probability of erroneously detecting a
significant difference. That is, the test statistic presented in
Eq~\ref{eq-w} has the expected type one error rate (or less in cases
where a small proportion of the population is sampled)
(Fig~\ref{fig-null}).

\begin{figure}

{\centering \includegraphics{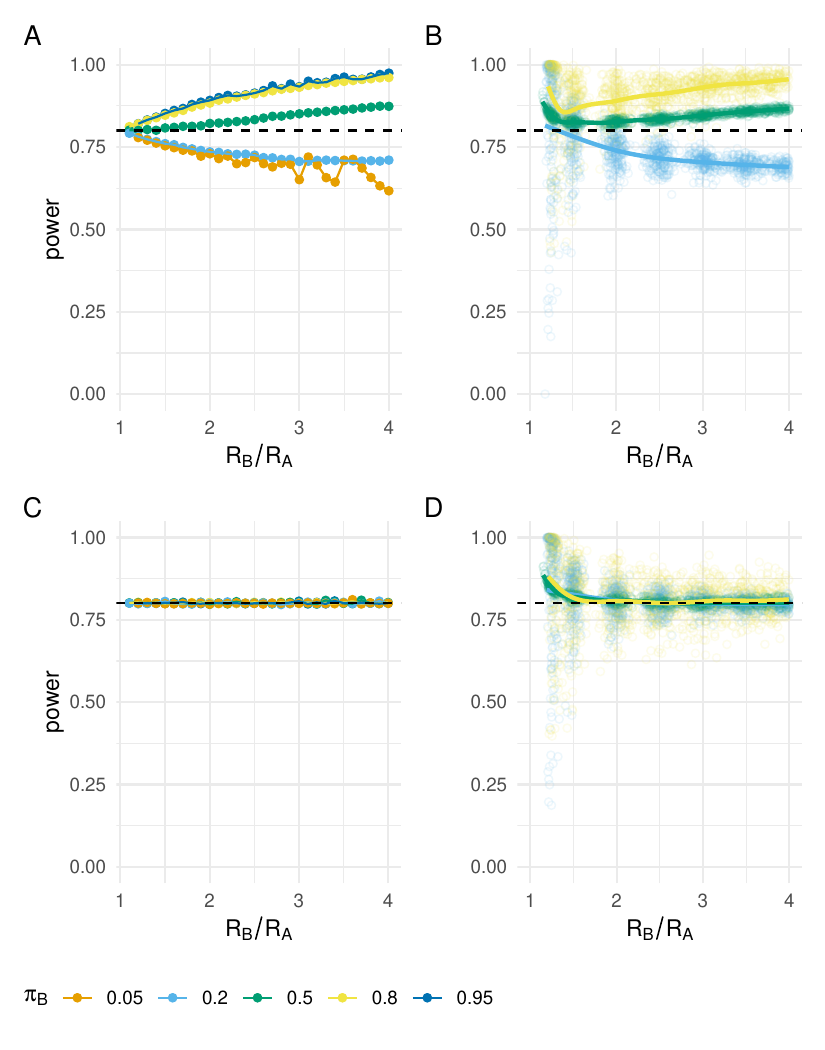}

}

\caption{\label{fig-alt}Simulated average power
(\(\alpha = 0.05, \beta= 0.2\), the nominal 80\% power is shown with the
dashed black line) varying the ratio of R values from 1 to 4 (x-axis)
and \(\pi_B = 0.05\) (orange), 0.2 (light blue), 0.5 (green), 0.8
(yellow), 0.95 (dark blue). A. Shows the theoretical simulation, B.
Shows the outbreak simulation C. Shows the theoretical simulation with
the `correction factor' applied. D. Shows the outbreak simulation with
the `correction factor' applied.}

\end{figure}

Selecting sample sizes based our our approach successfully identifies
significant differences in transmissibilty with the specified
probability. In the setting of perfect detection of linked pairs,
selecting sample sizes as per Eq~\ref{eq-samp} yields the expected power
in both theoretical and applied simulations (Fig~\ref{fig-alt} A,B).
That is, the actual power achieved is close to the target of 80\% when
the ratio of \(R_B/R_A\) is near one, and deviates as expected for a the
ratio of Poisson rates as per prior work by Ng and Tang (2005); with the
target power being (roughly) achieved when equal proportions of the
population are in each group, power being higher than targeted when the
more transmissible group is the majority, and lower than targeted when
the less transmissible group is the majority. We find that our
simulation based approach to correcting the sample size yields a study
population that achieves the desired power in all cases
(Fig~\ref{fig-alt} C,D).

\begin{figure}

{\centering \includegraphics{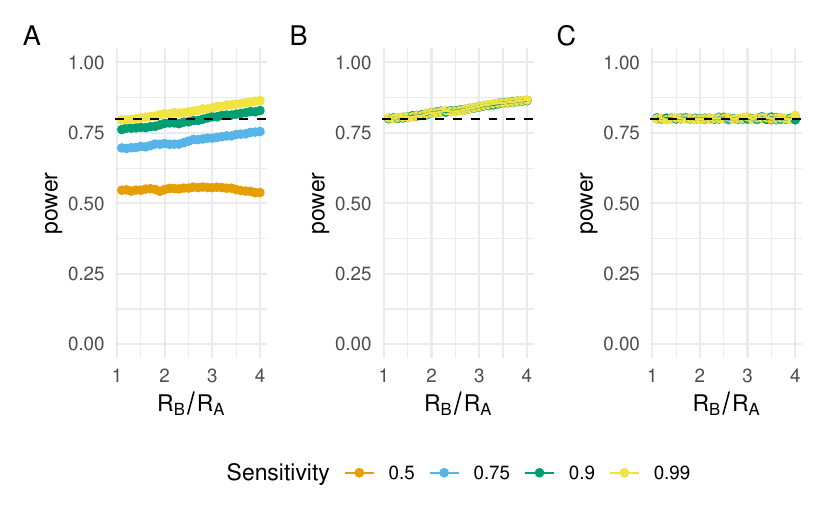}

}

\caption{\label{fig-sens-correction}Simulated average power
(\(\alpha = 0.05, \beta= 0.2\), the nominal 80\% power is shown in the
dashed black line) varying the ratio of R values from 1 to 4 (x-axis).
Here, the total number of infected individuals is 100,000 and the two
groups are of equal size. Each line shows a different sensitivity: 0.5
(orange), 0.75 (light blue), 0.9 (green), 0.99 (yellow). A. This panel
shows the impact on power if the sensitivity is not taken into account
in the sample size calculation. B. This panel shows the impact on power
when the sensitivity is taken into account in the sample size
calculation, demonstrating that the original power is recovered. C. This
panel shows the impact on power when the sensitivity is taken into
account and the `correction factor' is applied, bringing the power back
to 80\% as specified.}

\end{figure}

We examined the impact of imperfect detection of transmission linked
pairs (i.e., reduced sensitivity and specificity) on the power
calculations. Sensitivity is the probability that we will detect a
transmission linked pair that is in our sample. When this is less than
one, we will need a larger sample size than indicated by
Eq~\ref{eq-samp} (as confirmed by simulations,
Fig~\ref{fig-sens-correction}). Eq~\ref{eq-samp-sens} gives a corrected
sample size (Fig~\ref{fig-sens-correction} B), this can be further
optimized using our simulation-based approach
(Fig~\ref{fig-sens-correction} C).

\begin{figure}

{\centering \includegraphics{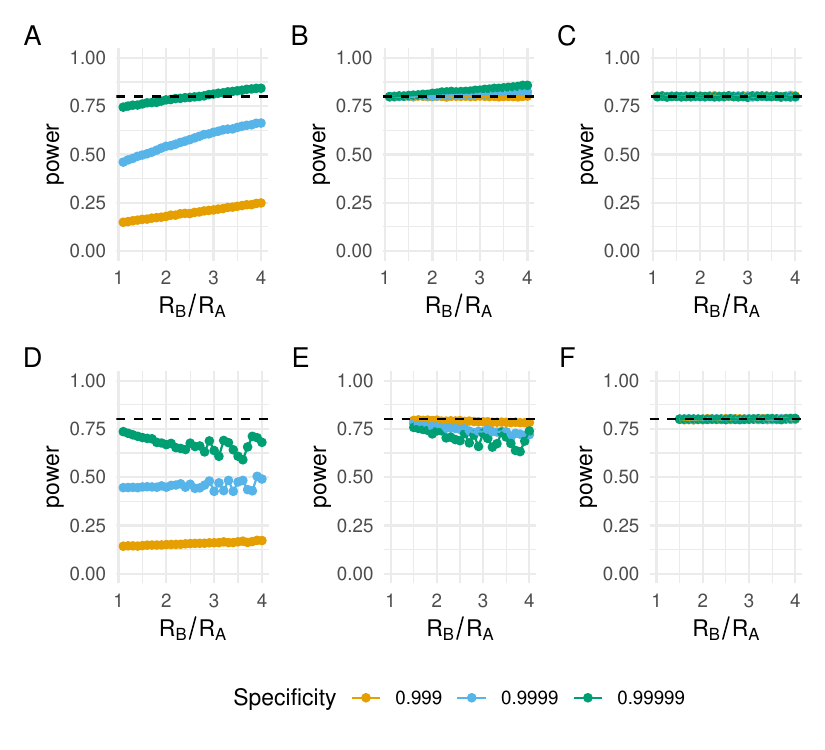}

}

\caption{\label{fig-spec-correction}Simulated average power
(\(\alpha = 0.05, \beta= 0.2\), the nominal 80\% power is shown in the
dashed black line) varying the ratio of R values from 1 to 4 (x-axis).
Here, the total number of infected individuals is 100,000. In the top
three panels (A, B, C) the two groups are of equal size; in the bottom
panels (D, E, F), \(\pi_B=0.05\), meaning 5\% of the infected population
are from group B, 95\% from group A. Each line shows a different
specificity: 0.999 (orange), 0.9999 (blue), 0.99999 (green) A. This
panel shows the impact on power if the specificity is not taken into
account in the sample size calculation. B. This panel shows the impact
on power when the specificity is taken into account in the sample size
calculation, demonstrating that the specified power is recovered,
although slightly over-powered, as is the case in general as the ratio
of R values increases. C. This panel shows the impact on power when the
specificity is taken into account in the sample size calculation and the
`correction factor' is applied, demonstrating that the specified power
is recovered, regardless of the ratio of R values. D. This panel shows
the impact on power if the specificity is not taken into account in the
sample size calculation. E. This panel shows the impact on power when
the specificity is taken into account in the sample size calculation,
demonstrating that the original power is not quite recovered, as is the
case when the groups are imbalanced with the smaller group in the
numerator (group B in this case). F. This panel shows the impact on
power when the specificity is taken into account in the sample size
calculation and the `correction factor' is applied, demonstrating that
the specified power is recovered, regardless of the ratio of R values.}

\end{figure}

Specificity is the probability that we will not link two people in our
sample who are in fact not a transmission linked pair. Perhaps more
intuitively, 1-specificity is the probability that two random people in
our sample would be considered a transmission linked pair if they are in
fact not linked. Because imperfect specificity biases estimates towards
the null (if errors are independent of the characteristic of interest),
imperfect specificity increases the needed sample size beyond that
indicated in Eq~\ref{eq-samp} (Fig~\ref{fig-spec-correction} A). This
corrected by using an iterative solving of Eq~\ref{eq-samp-sens-spec}
(Fig~\ref{fig-spec-correction} B) and further optimized by using our
simulation based approach (Fig~\ref{fig-spec-correction} C).

For examples of sample sizes given specific parameters, see Figures 1
and 2 in Appendix B.

\hypertarget{discussion}{%
\section{Discussion}\label{discussion}}

Here we have presented a simple estimator for the ratio of reproductive
numbers between two groups based on the identification of transmission
linked pairs, and provided validated tools for sample size calculation
based on this estimator. We have provided extensions to our method to
correct for imperfect identification of linked pairs, overdispersion in
the transmission process (i.e., the presence of super-spreading), and
group imbalance. In doing so, we aimed to provide accessible methods for
guiding study design, while retaining enough sophistication to deal with
the imperfect nature of molecular inferences about transmission. In
order to aid adoption, all methods are implemented in the
\texttt{phylosamp} R package available on CRAN and at
\url{https://github.com/HopkinsIDD/phylosamp} (Wohl \emph{and others}
2023). Use of this approach should lead to more efficient study designs
and better evaluation of information gleaned from molecular studies of
transmission, though more work is needed to address situations with more
complex dynamics (e.g., where transmission is assortative within
groups).

Because our approach is based only on linkage among individuals within
the same group, it is likely that it provides a conservative estimate of
the amount of information available in a given sample. If techniques or
data are available that allow us to leverage the information provided by
transmission pairs spanning groups, the sample size rendered by our
approach should be more than adequate to test the hypothesis of
differential transmission between groups (i.e., it should represent an
upper limit on the required sample size). For example, if the direction
of transmission could be known or inferred, linkages between individuals
with different characteristics could be better utilized. Likewise,
sophisticated techniques that take advantage of phylogenetic
relationships beyond direct linkage to characterize transmission
pathways may be able to more efficiently estimate the ratio of
reproductive numbers between groups, particularly if characteristics
represent distinct populations (e.g., different geographic regions);
thereby reducing the needed sample size. While more work is needed to
make studies with such additional information maximally efficient, use
of our method should ensure that they would be adequately powered.

Our assumption of transmission being independent of the relationship
between infector and infectee characteristics could be violated if there
is assortative mixing or if particular kinds of contacts were more
likely to lead to transmission than others. For example, different sex
acts have different probabilities of transmission for many STIs and are
more likely to occur in certain types of partnerships than others.
Extension of our approach to this setting is a clear avenue for future
work.

A further limitation of this approach is the assumption that sequences
come from a random sample of the infected population. This may be a
particular challenge for molecular studies as clinical or other
convenience samples are so often used. When the random sampling
assumption is violated, our methods may still be valid or correctable
using standard weighting techniques; doing so would require careful
consideration of the sampling process and potential biases.

However, it should be noted that our approach is robust to differential
susceptibility between groups. To understand why this is the case
consider that, absent assortative mixing, when we are looking at a
sample of only infected individuals we are unable to distinguish
differences in susceptibility between groups from differences in the
proportion of individuals in each group in the underlying population
(i.e., the ratio of individuals from each group in the infected
population is simply modified by the ratio of their susceptibility).
Since our estimator corrects for these proportions, it is not biased by
differential susceptibility between these populations. However, this
also means that it is impossible to distinguish differential
susceptibility between groups from differences in their prevalence in
the source population without information on uninfected individuals,
regardless of the sophistication of the techniques used.

As molecular data becomes increasingly important in the practice of
infectious disease epidemiology, it is important that we have accessible
and robust techniques to guide the collection of samples for a wide
variety of inferential goals. The approaches presented here are one
small step on the path to creating this set of tools. Though these
methods are limited to examining differential transmission between
people with different characteristics absent assortitivity, such focused
analyses are the key building blocks on the road to developing more
sophisticated techniques.

\hypertarget{references}{%
\section*{References}\label{references}}
\addcontentsline{toc}{section}{References}

\hypertarget{refs}{}
\begin{CSLReferences}{1}{0}
\leavevmode\vadjust pre{\hypertarget{ref-faria2014early}{}}%
Faria, N. R., Rambaut, A., Suchard, M. A., Baele, G., Bedford, T., Ward,
M. J., Tatem, A. J., Sousa, J. D., Arinaminpathy, N., Pépin, J., et al.
(2014). The early spread and epidemic ignition of HIV-1 in human
populations. \emph{science} \textbf{346}, 56--61.

\leavevmode\vadjust pre{\hypertarget{ref-fraser2009pandemic}{}}%
Fraser, C., Donnelly, C. A., Cauchemez, S., Hanage, W. P., Van Kerkhove,
M. D., Hollingsworth, T. D., Griffin, J., Baggaley, R. F., Jenkins, H.
E., Lyons, E. J., et al. (2009). Pandemic potential of a strain of
influenza a (H1N1): Early findings. \emph{science} \textbf{324},
1557--1561.

\leavevmode\vadjust pre{\hypertarget{ref-gardy2011whole}{}}%
Gardy, J. L., Johnston, J. C., Sui, S. J. H., Cook, V. J., Shah, L.,
Brodkin, E., Rempel, S., Moore, R., Zhao, Y., Holt, R., et al. (2011).
Whole-genome sequencing and social-network analysis of a tuberculosis
outbreak. \emph{New England Journal of Medicine} \textbf{364}, 730--739.

\leavevmode\vadjust pre{\hypertarget{ref-gu2008testing}{}}%
Gu, K., Ng, H. K. T., Tang, M. L. and Schucany, W. R. (2008). Testing
the ratio of two poisson rates. \emph{Biometrical Journal: Journal of
Mathematical Methods in Biosciences} \textbf{50}, 283--298.

\leavevmode\vadjust pre{\hypertarget{ref-lai2020early}{}}%
Lai, A., Bergna, A., Acciarri, C., Galli, M. and Zehender, G. (2020).
Early phylogenetic estimate of the effective reproduction number of
SARS-CoV-2. \emph{Journal of medical virology} \textbf{92}, 675--679.

\leavevmode\vadjust pre{\hypertarget{ref-macpherson2022unifying}{}}%
MacPherson, A., Louca, S., McLaughlin, A., Joy, J. B. and Pennell, M. W.
(2022). Unifying phylogenetic birth--death models in epidemiology and
macroevolution. \emph{Systematic biology} \textbf{71}, 172--189.

\leavevmode\vadjust pre{\hypertarget{ref-ng2005testing}{}}%
Ng, H. K. T. and Tang, M.-L. (2005). Testing the equality of two poisson
means using the rate ratio. \emph{Statistics in medicine} \textbf{24},
955--965.

\leavevmode\vadjust pre{\hypertarget{ref-ratmann2020quantifying}{}}%
Ratmann, O., Kagaayi, J., Hall, M., Golubchick, T., Kigozi, G., Xi, X.,
Wymant, C., Nakigozi, G., Abeler-Dörner, L., Bonsall, D., et al. (2020).
Quantifying HIV transmission flow between high-prevalence hotspots and
surrounding communities: A population-based study in rakai, uganda.
\emph{The Lancet HIV} \textbf{7}, e173--e183.

\leavevmode\vadjust pre{\hypertarget{ref-rose2019phylogenetic}{}}%
Rose, R., Hall, M., Redd, A. D., Lamers, S., Barbier, A. E., Porcella,
S. F., Hudelson, S. E., Piwowar-Manning, E., McCauley, M., Gamble, T.,
et al. (2019). Phylogenetic methods inconsistently predict the direction
of HIV transmission among heterosexual pairs in the HPTN 052 cohort.
\emph{The Journal of infectious diseases} \textbf{220}, 1406--1413.

\leavevmode\vadjust pre{\hypertarget{ref-stadler2012estimating}{}}%
Stadler, T., Kouyos, R., Wyl, V. von, Yerly, S., Böni, J., Bürgisser,
P., Klimkait, T., Joos, B., Rieder, P., Xie, D., et al. (2012).
Estimating the basic reproductive number from viral sequence data.
\emph{Molecular biology and evolution} \textbf{29}, 347--357.

\leavevmode\vadjust pre{\hypertarget{ref-volz2020}{}}%
Volz, E., Baguelin, M., Bhatia, S., Boonyasiri, A., Cori, A., Cucunuba
Perez, Z., Cuomo-Dannenburg, G., Donnelly, C., Dorigatti, I., Fitzjohn,
R., et al. (2020). Phylogenetic analysis of SARS-CoV-2. \emph{Imperial
College London}.
doi:\href{https://doi.org/10.25561/77169}{10.25561/77169}.

\leavevmode\vadjust pre{\hypertarget{ref-wohl2021sample}{}}%
Wohl, S., Giles, J. R. and Lessler, J. (2021). Sample size calculation
for phylogenetic case linkage. \emph{PLoS computational biology}
\textbf{17}, e1009182.

\leavevmode\vadjust pre{\hypertarget{ref-phylosamp}{}}%
Wohl, S., Lee, E. C., D'Agostio McGowan, L., Giles, J. R. and Lessler,
J. (2023). Phylosamp: Sample size calculations for molecular and
phylogenetic studies. \url{https://github.com/HopkinsIDD/phylosamp}.

\leavevmode\vadjust pre{\hypertarget{ref-wohl2020combining}{}}%
Wohl, S., Metsky, H. C., Schaffner, S. F., Piantadosi, A., Burns, M.,
Lewnard, J. A., Chak, B., Krasilnikova, L. A., Siddle, K. J., Matranga,
C. B., et al. (2020). Combining genomics and epidemiology to track mumps
virus transmission in the united states. \emph{PLOS biology}
\textbf{18}, e3000611.

\end{CSLReferences}

\end{document}


\maketitle
\ifdefined\Shaded\renewenvironment{Shaded}{\begin{tcolorbox}[frame hidden, breakable, interior hidden, borderline west={3pt}{0pt}{shadecolor}, sharp corners, boxrule=0pt, enhanced]}{\end{tcolorbox}}\fi

\hypertarget{appendix-a}{%
\section{Appendix A}\label{appendix-a}}

\hypertarget{distributional-assumptions}{%
\subsection{Distributional
Assumptions}\label{distributional-assumptions}}

Below is intuition behind our derivations. Let's begin by assuming we
are observing the whole outbreak of size \(N\). A transmission pair is
defined as two individuals \(i\) and \(j\) who are connected by an edge.
Assume \(e_{ij}\) is the probability that individual \(i\) infected
individual \(j\). The total number of people infected by individual
\(i\) is the sum, which we assume is Poisson distributed
\(e_{i\cdot}\sim Pois(R_{g_i})\), where \(R_{g_i}\) is the reproductive
number for individual \(i\) who has group level \(G=g\).

The total number of edges is equal to:

\[
E = \sum_{i = 1}^{N}e_{i.}
\] The sum of \(N\) Poisson distributed random variables is Poisson,
hence we can define the total number of edges in an outbreak as:

\[
E \sim Pois(NR_{pop})
\]

Assuming there are two groups \(g\in\{A,B\}\), the weighted average of
\(R_{g}\) across both groups, also known as \(R_{pop}\), can be written
as \(\pi_AR_A+\pi_BR_B\), where \(\pi_g\) is the proportion of the
outbreak belonging to group \(G=g\).

The total number of edges where a member of group \(G=g\) was the
infector can be written as:

\[
E_{g\cdot} = \sum_{\{i:G_i=g\}}e_{i\cdot}
\] The total number of individuals for whom \(g_i=g\) is \(N\pi_g\),
therefore \(E_{g\cdot}\sim Pois(N\pi_g R_g)\).

The total number of individuals in group \(G=g\) who are infected by
individual \(i\) is the sum of \(e_{ij}\) where \(G_j = g\) is:

\[
e_{i\cdot_{\{G_j=g\}}}=\sum_{j:G_j=g}e_{ij} \sim Pois(\pi_{g_j}R_{g_i})
\]

Since the proportion of the susceptible population belonging to group
\(G=g\), that is the group of the infectee, is \(\pi_{g_j}\).

The total number of edges where the infector and infectee are both
members of the same group, \(G=g\) is as follows:

\[
E_{gg} = \sum_{i:G_i=g}e_{i\cdot_{\{G_j=g\}}}
\]

The total number of individuals for whom \(G_i = g\) in an outbreak of
size \(N\) is \(N\pi_g\), therefore \(E_{gg}\sim Pois(N\pi_g^2R_g)\).

The above intuition assumes we are observing the entire outbreak. Under
a simple random sample, where \(M/N=\rho\) proportion of the outbreak
are sampled, the probability of sampling infector \(i\) is \(M/N\) and
the probability of sampling infector \(i\)'s infectee \(j\) is
\(\frac{M-1}{N-1}\). Therefore, the expected number of total number of
edges in a sample of size \(M\) from an outbreak of size \(N\) is:

\[
\mathbb{E}[E] = \frac{M(M-1)}{N-1}R_{pop}
\] where \(E\sim Pois\left(\frac{M(M-1)}{N-1} R_{pop}\right)\).
Similarly, the expected number of like edges, \(E_{gg}\) in a sample is:

\[
\mathbb{E}[E_{gg}]=\frac{M(M-1)}{N-1}\pi_g^2R_g
\] where \(E_{gg}\sim Pois\left(\frac{M(M-1)}{N-1} \pi_g^2R_g\right)\).

\hypertarget{sample-size-derivation}{%
\subsection{Sample size derivation}\label{sample-size-derivation}}

\begin{align*}
-z_{1-\beta}&=z_{1-\alpha}-\frac{\log\left(\frac{R_B}{R_A}\right)}{\sqrt{\left[1 + \frac{R_A}{d^2R_B}\right]\frac{(N-1)}{R_A\pi_A^2M(M-1)}}}\\
\sqrt{\left[1 + \frac{R_A}{d^2R_B}\right]\frac{(N-1)}{R_A\pi_A^2M(M-1)}} &= \frac{\log\left(\frac{R_B}{R_A}\right)}{z_{1-\alpha} + z_{1-\beta}} \\
\left[1 + \frac{R_A}{d^2R_B}\right]\frac{(N-1)}{R_A\pi_A^2M(M-1)} &= \frac{\log\left(\frac{R_B}{R_A}\right)^2}{(z_{1-\alpha} + z_{1-\beta})^2}\\
M(M-1) &= \left[1 + \frac{R_A}{d^2R_B}\right]\frac{(N-1)}{R_A\pi_A^2}\frac{(z_{1-\alpha} + z_{1-\beta})^2}{\log\left(\frac{R_B}{R_A}\right)^2}\\
M^2-M+\frac{1}{4} &= \left[1 + \frac{R_A}{d^2R_B}\right]\frac{(N-1)}{R_A\pi_A^2}\frac{(z_{1-\alpha} + z_{1-\beta})^2}{\log\left(\frac{R_B}{R_A}\right)^2} + \frac{1}{4}\\
\left(M - \frac{1}{2}\right)^2 &= \left[1 + \frac{R_A}{d^2R_B}\right]\frac{(N-1)}{R_A\pi_A^2}\frac{(z_{1-\alpha} + z_{1-\beta})^2}{\log\left(\frac{R_B}{R_A}\right)^2} + \frac{1}{4}\\
M -\frac{1}{2} &= \sqrt{\left[1 + \frac{R_A}{d^2R_B}\right]\frac{(N-1)}{R_A\pi_A^2}\frac{(z_{1-\alpha} + z_{1-\beta})^2}{\log\left(\frac{R_B}{R_A}\right)^2} + \frac{1}{4}}\\
M &= \sqrt{\left[1 + \frac{R_A}{d^2R_B}\right]\frac{(N-1)}{R_A\pi_A^2}\frac{(z_{1-\alpha} + z_{1-\beta})^2}{\log\left(\frac{R_B}{R_A}\right)^2} + \frac{1}{4}} + \frac{1}{2}\\
M &= \frac{1}{2}\left[\sqrt{4\left[1 + \frac{R_A}{d^2R_B}\right]\frac{(N-1)}{R_A\pi_A^2}\frac{(z_{1-\alpha} + z_{1-\beta})^2}{\log\left(\frac{R_B}{R_A}\right)^2} + 1} + 1\right]
\end{align*}

\hypertarget{appendix-b}{%
\section{Appendix B}\label{appendix-b}}

Below is code to estimate the sample size needed to estimate the ratio
between \(R_A: 0.83\), \(R_B: 1.67\), where the prevalence of the
characteristic of interest in the infected population is 0.8 for group A
(and therefore 0.2 for group B). We are targeting 80\% power for a
one-sided hypothesis test (testing \(H_0: R_B/R_A = 1\) vs
\(H_1: R_B / R_A > 1\)) with a significance level of \(\alpha = 0.05\).
The full outbreak size is 5,000.

\begin{Shaded}
\begin{Highlighting}[]
\FunctionTok{library}\NormalTok{(phylosamp)}
\FunctionTok{set.seed}\NormalTok{(}\DecValTok{7}\NormalTok{)}
\FunctionTok{relR\_samplesize}\NormalTok{(}\AttributeTok{R\_a =} \FloatTok{0.83}\NormalTok{,}
                \AttributeTok{R\_b =} \FloatTok{1.67}\NormalTok{,}
                \AttributeTok{p\_a =} \FloatTok{0.8}\NormalTok{,}
                \AttributeTok{N =} \DecValTok{5000}\NormalTok{,}
                \AttributeTok{alpha =} \FloatTok{0.05}\NormalTok{,}
                \AttributeTok{power =} \FloatTok{0.8}\NormalTok{,}
                \AttributeTok{alternative =} \StringTok{"greater"}\NormalTok{)}
\end{Highlighting}
\end{Shaded}

This estimates that we need a sample size of \texttt{1033} given the
above parameters. If instead of a one-sided hypothesis test we wanted to
conduct a two sided hypothesis test, we would run the following.

\begin{Shaded}
\begin{Highlighting}[]
\FunctionTok{relR\_samplesize}\NormalTok{(}\AttributeTok{R\_a =} \FloatTok{0.83}\NormalTok{,}
                \AttributeTok{R\_b =} \FloatTok{1.67}\NormalTok{,}
                \AttributeTok{p\_a =} \FloatTok{0.8}\NormalTok{,}
                \AttributeTok{N =} \DecValTok{5000}\NormalTok{,}
                \AttributeTok{alpha =} \FloatTok{0.05}\NormalTok{,}
                \AttributeTok{power =} \FloatTok{0.8}\NormalTok{,}
                \AttributeTok{alternative =} \StringTok{"two\_sided"}\NormalTok{)}
\end{Highlighting}
\end{Shaded}

A two-sided test with the above parameters is estimated to require a
sample size of \texttt{1164}. Since we have imbalanced groups (80\% are
in group A vs 20\% in group B) we know that this will be an
underestimate for the sample size needed, assuming that our
distributional assumptions are met. We can correct for this by using our
simulation-based approach by setting the parameter
\texttt{correct\_for\_imbalance\ =\ TRUE} as follows.

\begin{Shaded}
\begin{Highlighting}[]
\FunctionTok{relR\_samplesize}\NormalTok{(}\AttributeTok{R\_a =} \FloatTok{0.83}\NormalTok{,}
                \AttributeTok{R\_b =} \FloatTok{1.67}\NormalTok{,}
                \AttributeTok{p\_a =} \FloatTok{0.8}\NormalTok{,}
                \AttributeTok{N =} \DecValTok{5000}\NormalTok{,}
                \AttributeTok{alpha =} \FloatTok{0.05}\NormalTok{,}
                \AttributeTok{power =} \FloatTok{0.8}\NormalTok{,}
                \AttributeTok{alternative =} \StringTok{"two\_sided"}\NormalTok{,}
                \AttributeTok{correct\_for\_imbalance =} \ConstantTok{TRUE}\NormalTok{)}
\end{Highlighting}
\end{Shaded}

After correcting for the group imbalance using our simulation approach,
a two-sided test with the above parameters is estimated to require a
sample size of \texttt{1273}. Let's assume that we expect to imperfectly
observe linkage pairs such that the sensitivity is 0.9 and the
specificity is 0.999.

\begin{Shaded}
\begin{Highlighting}[]
\FunctionTok{relR\_samplesize}\NormalTok{(}\AttributeTok{R\_a =} \FloatTok{0.83}\NormalTok{,}
                \AttributeTok{R\_b =} \FloatTok{1.67}\NormalTok{,}
                \AttributeTok{p\_a =} \FloatTok{0.8}\NormalTok{,}
                \AttributeTok{N =} \DecValTok{5000}\NormalTok{,}
                \AttributeTok{alpha =} \FloatTok{0.05}\NormalTok{,}
                \AttributeTok{power =} \FloatTok{0.8}\NormalTok{,}
                \AttributeTok{alternative =} \StringTok{"two\_sided"}\NormalTok{,}
                \AttributeTok{sensitivity =} \FloatTok{0.9}\NormalTok{,}
                \AttributeTok{specificity =} \FloatTok{0.999}\NormalTok{,}
                \AttributeTok{correct\_for\_imbalance =} \ConstantTok{TRUE}\NormalTok{)}
\end{Highlighting}
\end{Shaded}

After incorporating sensitivity and specificity, our required sample
size increases to \texttt{2540}. Let's assume that there is
overdispersion present, with an overdispersion parameter of
\texttt{k\ =\ 35}. This overdispersion parameter implies that you are
assuming a negative binomial distribution as opposed to a Poisson
distribution, as is the default. An overdispersion parameter of infinity
would be equivalent to the default.

\begin{Shaded}
\begin{Highlighting}[]
\FunctionTok{relR\_samplesize}\NormalTok{(}\AttributeTok{R\_a =} \FloatTok{0.83}\NormalTok{,}
                \AttributeTok{R\_b =} \FloatTok{1.67}\NormalTok{,}
                \AttributeTok{p\_a =} \FloatTok{0.8}\NormalTok{,}
                \AttributeTok{N =} \DecValTok{5000}\NormalTok{,}
                \AttributeTok{alpha =} \FloatTok{0.05}\NormalTok{,}
                \AttributeTok{power =} \FloatTok{0.8}\NormalTok{,}
                \AttributeTok{alternative =} \StringTok{"two\_sided"}\NormalTok{,}
                \AttributeTok{correct\_for\_imbalance =} \ConstantTok{TRUE}\NormalTok{,}
                \AttributeTok{overdispersion =} \DecValTok{35}\NormalTok{)}
\end{Highlighting}
\end{Shaded}

After allowing for overdispersion, a two-sided test with the above
parameters is estimated to require a sample size of \texttt{4434}.

\hypertarget{estimated-sample-sizes}{%
\subsubsection{Estimated sample sizes}\label{estimated-sample-sizes}}

The following figures demonstrate the estimated sample sizes needed
given specific parameters.

\begin{figure}

{\centering \includegraphics{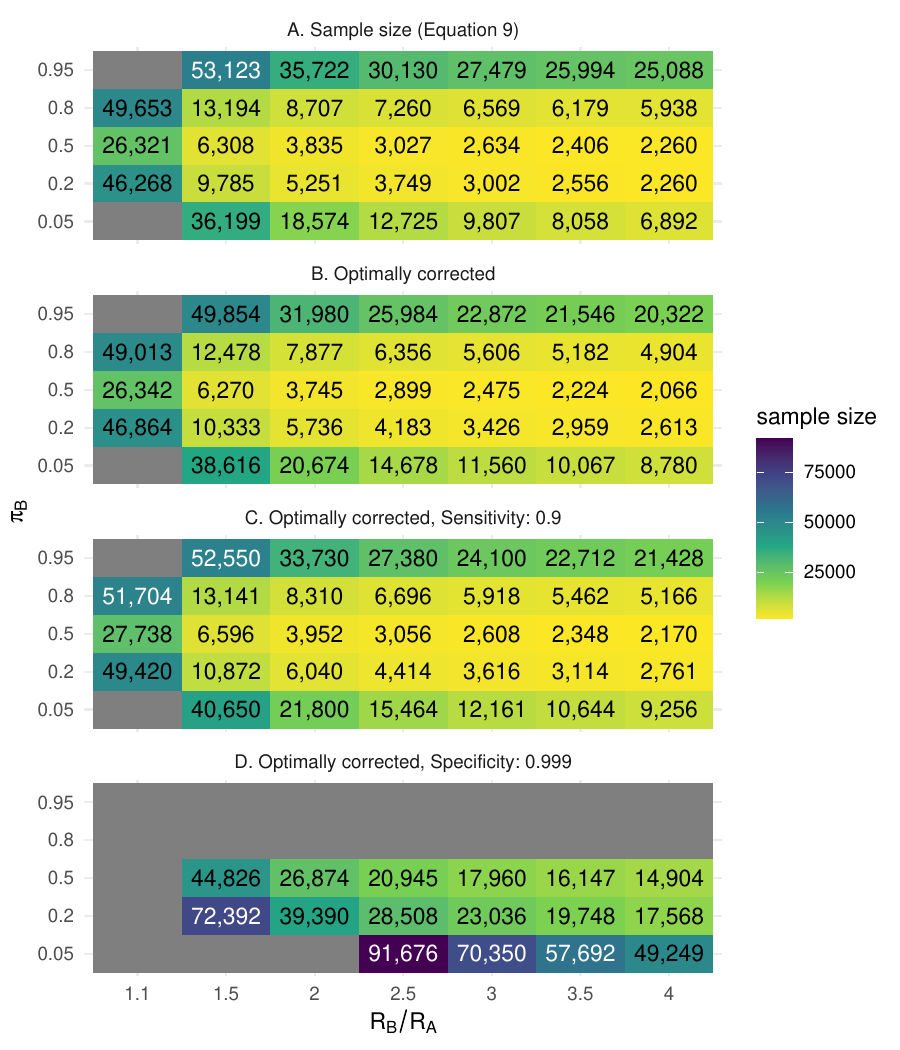}

}

\caption{The sample size needed for an epidemic of size 100,000 with
\(\pi_B\) as specified on the y-axis and the ratio between \(R_B\) and
\(R_A\) as specified on the y-axis, chosen such that
\(R_A\pi_A + R_B\pi_B = 1\) to achieve 80\% power for a one-sided test
where \(\alpha = 0.05\). The top two panels (A, B) correspond to a
perfectly observed sample (i.e.~no sensitivity or specificity). Panel C
has a sensitivity of 0.9, the bottom panel (D) has a specificity of
0.999. Panel A uses the sample size calculation given in Equation 9.
Panels B-D use the simulation-based optimal sample size, using methods
in the \texttt{phylosamp} R package. Grey boxes indicate parameter
combinations for which the sample size needed to detect the requested
change exceeds the size of the epidemic.}

\end{figure}

\begin{figure}

{\centering \includegraphics{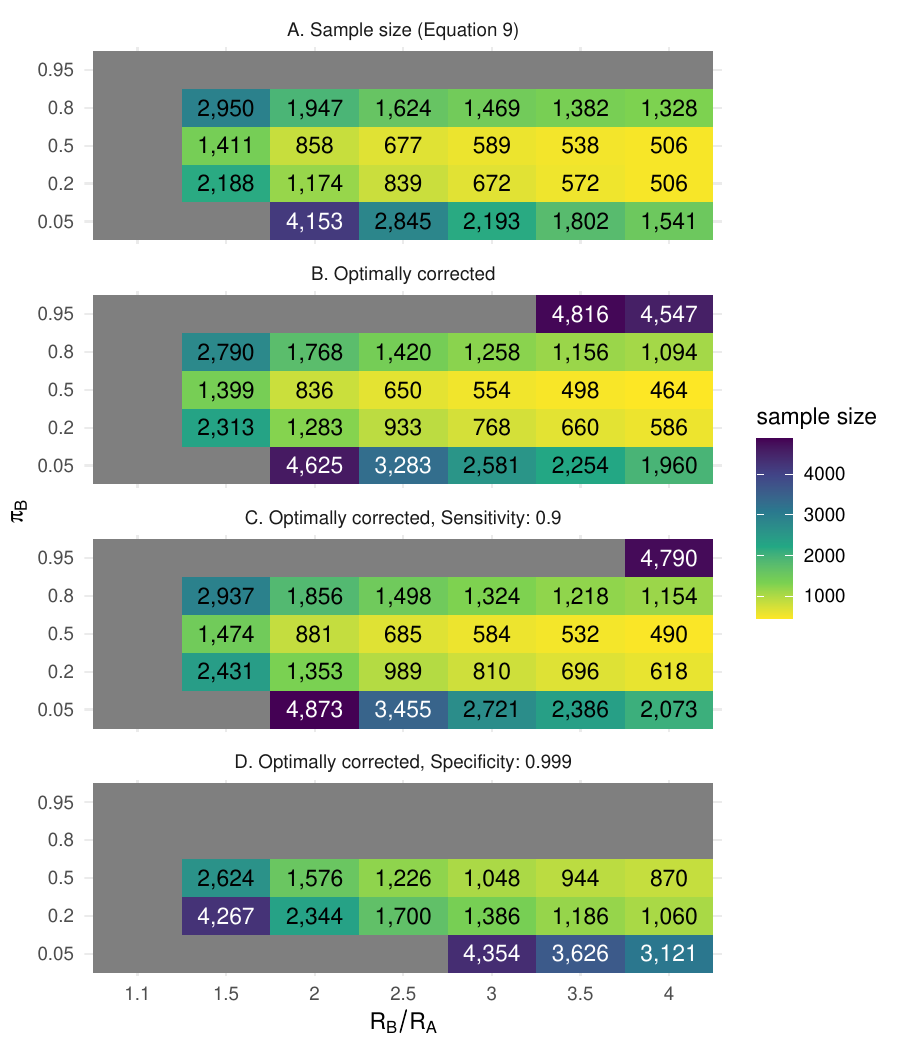}

}

\caption{The sample size needed for an outbreak of size 5,000 with
\(\pi_B\) as specified on the y-axis and the ratio between \(R_B\) and
\(R_A\) as specified on the y-axis, chosen such that
\(R_A\pi_A + R_B \pi_B= 1\) to achieve 80\% power where
\(\alpha = 0.05\) . The top two panels (A, B) correspond to a perfectly
observed sample (i.e.~no sensitivity or specificity). Panel C has a
sensitivity of 0.9, the bottom panel (D) has a specificity of 0.999.
Panel A uses the sample size calculation given in Equation 9. Panels B-D
use the simulation-based optimal sample size, using methods in the
\texttt{phylosamp} R package. Grey boxes indicate parameter combinations
for which the sample size needed to detect the requested change exceeds
the size of the outbreak.}

\end{figure}